# Toward an Interactive Directory for Norfolk, Nebraska: 1899-1900

**Robert B. Allen**
Research Center for Knowledge Communities
University Tsukuba
Tsukuba, Ibaraki, Japan
rba@boballen.info

**Abstract:**
*We describe steps toward an interactive directory for the town of Norfolk, Nebraska for the years 1899 and 1900. This directory would extend the traditional city directory by including a wider range of entities being described, much richer information about the entities mentioned and linkages to mentions of the entities in material such as digitized historical newspapers. Such a directory would be useful to readers who browse the historical newspapers by providing structured summaries of the entities mentioned. We describe the occurrence of entities in two years of the Norfolk Weekly News, focusing on several individuals to better understand the types of information which can be gleaned from historical newspapers and other historical materials. We also describe a prototype program which coordinates information about entities from the traditional city directories, the federal census, and from newspapers. We discuss the structured coding for these entities, noting that richer coding would increasingly include descriptions of events and scenarios. We propose that rich content about individuals and communities could eventually be modeled with agents and woven into historical narratives.*

## 1 INTRODUCTION

Newspapers provide a particularly rich foundation for organizing other types of historical and cultural resources about a community. Historical newspapers have a wide range of rich content about the communities they cover. A great many historical newspapers have been digitized, OCRd, and access provided with basic search interfaces. A range of user interfaces might be developed to support access to that digitized material. We describe some approaches to systematically describing the content in ways that might be useful for advanced interfaces.

Extracting the rich information in historical newspapers and organizing it in a useful form is a great challenge. The organizational methods must be able to describe the relationships among the entities and events which are found in the material. Ultimately, the organization will need to be "model-oriented" (Allen, 2013a, b). Because the material is so complex, developing a model-oriented view is major undertaking. Here, we describe a first step in that direction by focusing on the entity instances which appear in a newspaper for a small Midwestern U.S. town during the years 1899 to 1900.

The entries in historical materials are much more complex than those in a typical database. Consider the details we might want to record about an individual. They include relatively straightforward attributes such as gender, age, and ethnicity, more complex attributes such as education and health, and highly detailed situation-specific attributes such as actions at a specific time and place (see Allen, 2013a, b). General notions such as "linked data" are too vague for this application. The key factor is the type of structure adopted for describing the nuances.

Complex entity structures need to be developed for people and for organizations. The entity structures for people could be the foundation of biographies. Entity structures for organizations could be similar to the knowledgebase of the structure of the federal government described by (Allen, 2011b). Such structures should improve the accuracy of text mining and provide focus for the most important types of material to be mined.

These models should also allow coordination across different types of content at the level of descriptions of entities and events. For instance photographs, newspapers, church records, and even physical artifacts could be connected by descriptions. We have described a case of coordination across two newspapers from the same town (Allen, 2010). Our approach to information organization could form a framework for cultural collections such as Europeana and the Digital Public Library of America (DPLA).

As organizing digital library collections proceeds, standards are needed to promote consistency. Ultimately, the emphasis will be on a unified description of the community in which individuals and institutions are coordinated. We term this broader perspective a "community model".

City directories were common in towns across the United States from the 1870s until they were supplanted by telephone directories in the early 20$^{th}$ century. While there was variation across publishers, the city directories typically included addresses and occupations for established residents of the towns. They also included listings of businesses and usually advertisements for businesses.

In this paper, we describe the issues in developing an interactive city directory for the town of Norfolk, Nebraska. Potentially, the interactive city directory would enable users to obtain a far richer understanding of the individuals and organizations included in that directory than is possible with traditional paper directories. Moreover, the directories could serve as a platform for the development of more complex interactive descriptions of the town. We close by discussing some of the implications of this approach, such as to the study of history and models of communities.

## 2   A SPECIFIC EXAMPLE: NORFOLK, NEBRASKA

### 2.1   Selecting Norfolk, Nebraska

For this study we sought a town which fit certain criteria. The town needed to have a digitized newspaper with an OCR transcription. And, it needed to be of small size and relatively self-contained so as to limit the complexity of the contents of the newspaper. In addition, the full text of supplemental materials also needed to be available. We were particularly interested in the availability of transcribed census materials. It was also helpful if other materials such as transcriptions of existing traditional city directories, diaries, and oral histories were available.

Based on these criteria, we selected Norfolk, Nebraska, the largest town in Madison County. The *Norfolk Weekly News* was available from the LC/NEH National Digital Newspaper Program (NDNP) in the United States for the years 1899 to 1900. Limited transcriptions for the City Directory for 1889 and for a Business Directory for 1899 were available at the Madison Country Genealogical Society website[1]. Moreover, FamilySearch.org kindly provided a transcription of the 1900 census for Madison County, Nebraska.

Norfolk was prosperous and rapidly growing in this time period. It was the major town in a fertile farming region and the fifth largest town in Nebraska. Sugar beets were the main crop, there was a sugar-refining plant, and the town was nicknamed "Sugar City". It was a railroad junction with connections to the larger towns of Sioux City, Iowa toward the east, Omaha, Nebraska to the south, and the farming territory of northern Nebraska to the west.

Norfolk was served by two newspapers -- the *Norfolk Weekly News*, which has been digitized and is available through the NDNP, and the *Norfolk Times Tribune*, which is available on microfilm but has not been digitized. Typically, the *Norfolk Weekly News* included two pages of national and international news followed by a few pages of regional news and lifestyle guidance. This was usually followed with several pages of brief notices about events in the town. These notices ranged from the roster of teachers by classroom in the school, to itemized payments approved by the town government, listings of visitors from out of town or travels by the town's residents, to property transfers and shop openings.

The transcription of the 1900 census for Madison County has 17,621 entries. For the town of Norfolk and immediately surrounding township there were 5,176 entries. The census lists attributes such as age, race, ethnicity, and place of birth. In addition, it is arranged by household with the husband listed as head of household. The region was also the home to a psychiatric hospital and the patients there are included in the census, but the hospital is rarely mentioned in the newspaper and its patients were not included in our analyses.

While these are particularly rich sources of information, some caution is needed in using the data. In particular, there are many errors in the OCR transcriptions. While most names appeared to have been recognized, it is possible that particularly unusual names were incorrectly identified. For instance, the name McClary[2] was sometimes confused with the last letters of the word "secretary".

---

[1] http://mcgs.nesgs.org/
[2]  J.S. McClary was a Civil War veteran and judge. His son was a census taker for the 1900 census, some of which is used here, and Edith McClary was school teacher.

## 2.2 General Observations from the *Norfolk Weekly News*

The time period on which we focus is a distinctive era. The economy of the Midwest was growing as railroads opened new territories and markets. Print technology was adequate for relatively widely distributed newspapers that retained a regional emphasis.

The *Norfolk Weekly News* reported local residents' travels. While we expected some travel, we had presumed that the town would be a relatively closed system. In fact, there was a constant flow of visitors to the town and residents of the town often traveled to other towns and frequently to nearby states. In effect, "the community" covered by the *Norfolk Weekly News* extended across a large area and there is discussion in the newspaper of towns with similar concerns several states away.

Overall, Norfolk appeared to be a relatively stable and homogeneous community. It was founded by German-American settlers and many of the citizens were identified in the census as ethnically of German extraction. When race was identified it was always listed as "White". However, news reports refer to a "Norfolk colored club" or "Colored Lime Kiln" that held regular celebrations for Emancipation Day. Two individuals were identified in the census as having been born in China and one of them, "Sam Goon" was identified as the head of a family. Sam Goon appears once in the newspaper as having been paid $1.00 for "labor on the streets". One head of household was listed in the census as having been born in Asia Minor. It also seems likely that there would be some people of Native American ancestry in the community but none were found from this preliminary investigation.

## 2.3 Coverage in Newspaper of Individuals Identified in Census

We selected the names of 10 people at random from the 1900 census list for Norfolk and the immediately surrounding area. The occurrence of their names in newspapers for 1899 and 1900 was checked. Because the children and elderly citizens selected in the random sample were rarely mentioned in newspapers, we considered anyone from the households of which those individuals were members.[3] References to members of six of the 10 families were found in the two years of the newspaper studied. Thus, it appears that the majority of the town's population was covered by the newspaper but also a substantial number of members of the community were not represented.

## 2.4 High-Profile Individuals

Some individuals and their families were mentioned frequently in the newspaper. To the extent that this project is considered as developing a guide to the newspaper (rather than as creating a balanced or comprehensive model of the community), then it useful to consider those individuals and families which appear most frequently.

### 2.4.1 Carl Asmus and Family

Carl Asmus ran the principal grocery in the town. His family was active and their activities were frequently reported. Overall, there were 22 mentions of the family name in the two year period. For instance his son, Max Asmus, took long bicycle trips that were frequently reported in the newspaper.

---

[3] An exception was a one-year old child who was selected as part of the sample and then was reported in the newspaper as having died of diphtheria.

### 2.4.2 Henry Baird

Mr. Baird was the manager at the sugar factory. His name appeared 17 times in the two years of newspapers studied. He is mentioned as having been involved in many social activities in 1899, was married in 1900, and then moved to Colorado with his bride.

### 2.4.3 J.E. Simpson

Mayor Simpson, who was also an insurance agent, was mentioned 171 times in the two-year period studied. About two-thirds of those references were to his role as mayor. For example, he chaired Council meetings and made remarks at several public occasions.

### 2.4.4 H.C. Truman

H.C. Truman owned a paint-and-wallpaper shop and also had a painting business which advertised regularly in the newspaper. Overall, the Truman name appeared about 81 times in the two-year period. He was also mentioned frequently as being paid by the government for paint jobs. He was particularly active in ward-level politics and the Fireman's Association.

## 3 PROGRAM IMPLEMENTATION

We developed a Java language computer program for extracting individuals from the census and providing a class instance for each individual and also classes for households. The program also included Businesses and Buildings as entities. The Government structure was similar to that from (Allen, 2011b). Moreover, the program allows interactive queries about the entities. Thus, it is a prototype for the interactive extended city directory. A flexible structure was implemented for adding new entity instances and details about existing entities. Typically, that additional information would be gleaned from the newspapers. That knowledgebase is still to be richly populated.

## 4 IMPLICATIONS AND APPLICATIONS

### 4.1 Supporting Interaction with Historical Newspapers

Newspapers are particularly complex information resources and there are many possible ways of organizing their contents (e.g., Allen & Hall, 2010; Liddy et al., 1993). We have previously proposed that timelines may support interaction with historical newspapers (e.g., Allen, 2011a) and we propose here that an interactive directory may be helpful. We envision a set of increasingly nuanced directories which could be coordinated with timelines. Different interfaces will be needed for users with different information needs and user studies will help determine the most useful interfaces for each of those types of information needs. We can provide tools to support individual historians (Allen & Sieczkiewicz, 2010) and collaboration among groups of scholars (Shaw, 2013).

### 4.2 Role of the Newspaper in the Community

A distinctive aspect of Allen's (2013a, b) approach to description is that it provides a mechanism for incorporating newspapers as artifacts of the community rather than as authoritative reports of the community. Our models would incorporate the creation and use of newspapers. This is analogous to the point made by Upward (1997) with respect to archives. Just as Upward argued that archives both reflect and provide structure to society through structuration (Giddens, 1984), newspapers may also be seen as contributing to "structuration" of the community.

## 4.3 Inclusion of, and Extension to, Other Types of Content

It was surprisingly difficult to find towns which met our relatively modest criteria. Fortunately, in the case of Norfolk, there are a great many additional resources from this time period which could be incorporated. We suggest that the ability to coordinate with other digitized resources be considered a major factor in the selection of future digitization projects. As more resources are digitized, we should be able to find more overlaps.

Many images are available for Norfolk (see Schmeckpeper, 2004 for a sample). Other post cards, including some available on the Web, include handwritten notes which could be incorporated into the records of the community. Many family genealogies are now available and some of those include family photographs. Search-engine searches may facilitate coordination of the varied resources. For example, a search shows that there are many people named Asmus (see Section 2.4.1) still in the Norfolk area. As a way of coordinating these resources, the present approach could be seen as developing multi-biographies or multi-family genealogies. Essentially, we envision a new medium for the organization and presentation of local history.

## 4.4 Beyond an Entity-Based Directory

We can collect much information about entities but the information is often so rich that simple descriptions of entities cannot handle the complexity. For instance, while people are entities in this approach, people change over time and those changes are very hard to handle with simple structures. Moreover, the changes are often threaded together with connected events (Allen, 2013a) and those events often connected in stories associated with individuals. For instance, the newspaper reported that a 24-year old baker and a 14-year old girl eloped, and were caught. The mother of the girl objected to the marriage but then relented and gave the required legal consent[4]. More abstractly, the services and processes typically offered by businesses could be described with strings of events.

A consistent framework for coordinating events is needed. This linking of entities and events could be considered a type of linked data but is better-characterized as "model-oriented". Allen (2013a) suggested the use of verb frames based on the FrameNet projects to develop these models. Additional formalisms might be adopted from software engineering and artificial intelligence such as object-oriented modeling, conceptual modeling for requirements, frames, project management, service science, blackboard systems, software modularity, and temporal databases. In general, we embrace the rich structures these approaches support, but we deemphasize the role of inference because of the complexity of the material with which we are dealing. Repeating the inferences and explanations made by humans with citation to the source seems wiser than allowing complex automatic inferences.

## 4.5 Toward Structured Biographies and Community Models

While we have focused on one town and there is more data about it than we can readily handle with current tools, we envision that the development of better tools will allow not only complete processing of Norfolk but also extension to other towns. There would be many synergies from a much larger collection. Both people and events could be cross referenced from the perspective of different resources.[5]

---

[4] *Norfolk Weekly News*, October 26, 1899, Page 10.
http://chroniclingamerica.loc.gov/lccn/sn95070060/1899-10-26/ed-1/seq-12.pdf
[5] For instance, an oral history from the "Documenting the American South" project has a brief mention of Norfolk Nebraska. http://docsouth.unc.edu/sohp/I-0083/excerpts/excerpt_852.html

Complex structures, rather than simple ontologies, are needed for organizing this material. A highly detailed structure will be needed for describing individuals; this structure might be mined from interrelationships among materials on the Web. In some cases, we may have diaries, interviews, and oral histories on which to draw. The text of these materials could be cast into natural-language based frames (see Allen 2013a). They may also be used to develop structured representations for biographies (Allen 2013a, b). Because people evolve over time, it may be easier to generate descriptions of them with program-like scripts. This is especially true for descriptions of transient states such as emotions.

Ideally, biographical models would also interoperate with other types of models such as household models and community models. We have suggested that the representation of entities and events for a town could be a model of the community (Allen et al., 2007). By modeling the interaction of many people in many different ways, we could develop historical social networks. Community models would also allow comparisons among different communities. For instance, we might compare farming communities with waterfront fishing communities or manufacturing communities. Evaluation of those comparisons remains a significant challenge. On one hand, Norfolk seems typical of many small towns. On the other hand, the mere fact that it has a newspaper which reports events in such detail is unusual and it may be atypical in other ways.

Much of what defines a community are standard processes. These may be rituals, procedures, proceedings, agendas, or other sorts of workflows. They need to be documented and saved. Because such workflows are defined with different levels of detail, variability needs to be incorporated into the models. Moreover, unlike most workflows employed in computer and information science, many community, social, and cultural flows may be interrupted, side-tracked, or abandoned. For instance, we may describe the 30-year long effort to build a railroad line from Norfolk about 40 miles to Yankton, South Dakota. This would include specific plans for the path of the railroad, progress in obtaining financing, and initial steps in construction. But, it would also include the failure of the first attempt, regrouping and discussion of re-launching the project, and eventual abandonment of the effort. Ultimately, the description of these processes could include a discussion of some of the reasons for the failure of the project such as the challenge that the Norfolk-Yankton route posed for the existing rail lines through Sioux City.

Even more ambitiously, we might explore the potential to incorporate "computational sociology". Sociological theories, especially structuralist theories, might guide the selection of representations for entities and events. Structuralist theories from anthropology might be incorporated also into the community models. "Grand theories" such as those of Parsons (1937)[6] provide potential frameworks for describing the stability and coherence of open social systems that may also be useful in organizing the cultural materials.[7]

While there are many events described in the *Norfolk Weekly News,* there are even more gaps. Even the most prominent citizens are mentioned only about once a week on average. Some level of inference is inevitable. While some inferences seem natural (e.g., that a person eats and sleeps over several days) others must be made with extreme caution. To an extent, the plausibility of assumptions can be evaluated based on whether the simulations match trusted observations. But, as with inferences made in narrative history, these are likely to be controversial.

---

[6] Whether these sociological theories explain social change as well as stability is an ongoing controversy.

[7] Parsons is sometimes described as a "structure-functionalist". He proposes a general systems description of society as requiring four interacting components, and associates specific functions to those components which could be modeled.

### 4.6 Implications for the Study of History

This is a "big data" approach to history that is also focused on low-level details about individuals and communities. There are many ways the techniques could be applied on a large scale across towns. Robust text-extraction programs will help match the entities and attributes of the models but will likely encounter many challenges (see Crane & Jones, 2005). It is likely that a crowd-sourced solution[8] will be needed to provide human judgments for some cases.

Some historians (e.g., Short & Bradley, 2004) have explored "prosopography" as an approach to studying groups of historical persons. Our approach complements the prosopographic approach in the context of a well-documented town with detailed information about a significant portion of individuals. Richly interlocking information about individuals, about the social context of those individuals, and about the community as a whole can be modeled when we have rich data sources such as historical newspapers.

Simulations might be used for multi-agent prosopography in which the dynamics of social interactions would be modeled. They may support engaging interactive education and perhaps history-based games or even cyber-dramas. A personalized history game might even allow interaction with avatars based on one's own ancestors.

## 5  CONCLUSION

This has been an exploratory study to determine the potential for building an interactive directory for the town of Norfolk, Nebraska. Such an interface is possible and would be increasingly useful, after developments at many levels.

## 6  ACKNOWLEDGEMENTS

I thank Family Search for providing the transcription of the 1900 census records for Madison County, Nebraska. I also thank the Madison County (Nebraska) Genealogical Society for posting transcriptions of several historical documents relating to Norfolk, Nebraska.## 7  REFERENCES

Allen, R.B., Improving Access to Digitized Historical Newspapers with Text Mining, Coordinated Models, and Formative User Interface Design, *IFLA International Newspaper Conference*: Digital Preservation and Access to News and Views, 2010, 54-59. http://boballen.info/RBA/PAPERS/IFLA2010/iflaDelhi.pdf
Allen, R.B., Visualization, Causation, and History, *iConference*, 2011a, doi: 10.1145/1940761.1940835
Allen, R.B., Developing a Knowledge-base to Improve Interaction with Collections of Historical Newspapers, *IFLA WLIC,* August 2011b, San Juan, PR. http://conference.ifla.org/past/ifla77/188-allen-en.pdf
Allen, R.B., Model-Oriented Information Organization: Part 1), *D-Lib Magazine*, July, 2013, doi: 10.1045/july2013-allen-pt1
Allen, R.B., Model-Oriented Information Organization (Part 2), *D-Lib Magazine*, July, 2013, doi: 10.1045/july2013-allen-pt2
Allen, R.B., Japzon, A., Achananuparp, P., and Lee, K-J., A Framework for Text Processing and Supporting Access to Collections of Digitized Historical Newspapers. *HCI International Conference*, 2007, doi: 10.1007/978-3-540-73354-6_26---

[8] Zarndt, F., Putting the World's Cultural Heritage Online with Crowd-Sourcing, *IFLA Newspaper Section Satellite Meeting,* 2012, Mikkeli Finland, http://www.ifla2012mikkeli.com/getfile.php?file=125